\documentclass[preprint,superscriptaddress,showpacs,preprintnumbers,amsmath,amssymb]{revtex4}

\usepackage{booktabs}
\usepackage{graphicx,color}
\usepackage{dcolumn}
\usepackage{bm}
\usepackage{CJK}
\usepackage{mathrsfs}
\usepackage[dvipdfm,
            pdfstartview=FitH,
            CJKbookmarks=true,
            bookmarksnumbered=true,
            bookmarksopen=true,
            colorlinks,
            pdfborder=001,
            linkcolor=blue,
            anchorcolor=blue,
            citecolor=blue
            ]{hyperref}

\begin{document}
\begin{CJK}{GBK}{song}

\title{Relativistic Brueckner-Hartree-Fock theory for neutron drops}

\author{Shihang Shen}
\affiliation{State Key Laboratory of Nuclear Physics and Technology, School of Physics,
Peking University, Beijing 100871, China}
\affiliation{ Department of Physics, The University of Hong Kong, Pokfulam Road, Hong Kong, China}

\author{Haozhao Liang}
 \affiliation{RIKEN Nishina Center, Wako 351-0198, Japan}
 \affiliation{Department of Physics, Graduate School of Science, The University of Tokyo, Tokyo 113-0033, Japan}
%
\author{Jie Meng\footnote{Email: mengj@pku.edu.cn}}
 \affiliation{State Key Laboratory of Nuclear Physics and Technology, School of Physics,
Peking University, Beijing 100871, China}
 \affiliation{Department of Physics, University of Stellenbosch, Stellenbosch, South Africa}
 \affiliation{Yukawa Institute for Theoretical Physics, Kyoto University, Kyoto 606-8502, Japan}
\author{Peter Ring}
 \affiliation{State Key Laboratory of Nuclear Physics and Technology, School of Physics,
Peking University, Beijing 100871, China}
 \affiliation{Physik-Department der Technischen Universit\"at M\"unchen, D-85748 Garching, Germany}
%
\author{Shuangquan Zhang}
 \affiliation{State Key Laboratory of Nuclear Physics and Technology, School of Physics,
Peking University, Beijing 100871, China}

\date{\today}

\begin{abstract}
Neutron drops confined in an external field are studied in the framework of relativistic Brueckner-Hartree-Fock theory using the bare nucleon-nucleon interaction. The ground state energies and radii of neutron drops with even numbers from $N = 4$ to $N=50$ are calculated and compared with results obtained from other nonrelativistic \textit{ab initio} calculations and from relativistic density functional theory. Special attention has been paid to the magic numbers and to the sub-shell closures. The single-particle energies are investigated and the monopole effect of the tensor force on the evolutions of the spin-orbit and the pseudospin-orbit splittings is discussed. The results provide interesting insight of neutron rich systems and can form an important guide for future density functionals.
\end{abstract}


\maketitle

\section{Introduction}

The shell structure in atomic nuclei is one of the most astonishing facts. It has been discovered in the late 1940s \cite{GoeppertMayer1949_PR75-1969,Haxel1949_PR75-1766} and forms the foundation of nuclear physics since then.
With the advance of radioactive ion beam facilities around the world, more and more neutron-rich exotic nuclei have been reached, where many interesting new phenomena emerge such as the disappearance of traditional and the appearance of new magic numbers~\cite{Bastin2007_PRL99-022503,Sorlin2008_PPNP61-602,Steppenbeck2013_Nature502-207}, the halo phenomena~\cite{Tanihata1985,Meng1996,Meng1998}.
These new findings present challenges for current nuclear structure theory, in particular for nuclear density functional theory which should provide a uniform description over the entire nuclear chart~\cite{Vautherin1972_PRC5-626,Walecka1974_APNY83-491,Decharge1980_PRC21-1568,Bender2003_RMP75-121,Bogner2013_CPC184-2235}. In this framework there is no possibility to adjust the effective interaction or the resulting single particle energies separately for each area of the chart, as it is done in many shell model configuration interaction (CI) calculations.
Nuclear density functionals, even though they are phenomenological, are usually obtained by fitting to the properties of stable nuclei and, therefore, they are not well constrained in exotic regions far from the line of $\beta$-stability.
Microscopic calculations started from nucleon-nucleon ($NN$) interaction, or the so called \emph{ab initio} calculations \cite{Day1967_RMP39-719,Dickhoff2004_PPNP52-377, LEE-D2009_PPNP63-117,LIU-Lang2012_PRC86-014302,Barrett2013_PPNP69-131,Hagen2014_RPP77-096302,Carlson2015_RMP87-1067, Hergert2016_PR621-165,SHEN-SH2016_ChPL33-102103,Shen2017}, can provide valuable information to understand nuclear structure but are still difficult to be applied for exotic nuclei.

The neutron drop provides an ideal simple model to investigate the neutron-rich environment.
It is composed of finite number of pure neutrons which are constrained in an external field to keep the neutrons bound.
Because only the neutron-neutron interaction exists, equations for neutron drops are much easier to be solved. Therefore they can be calculated by many \emph{ab initio} methods \cite{Pudliner1996_PRL76-2416,Bogner2011_PRC84-044306,Maris2013_PRC87-054318,Potter2014_PLB739-445,Shen2018}.
In this way, different methods and different interactions can be compared, and valuable information can be obtained for constraining nuclear density functionals in neutron-rich system.

The neutron drops were first studied by quantum Monte Carlo methods~\cite{Pudliner1996_PRL76-2416} for $N = 7$ and $8$ using the two-nucleon (2N) interaction Argonne $v_{18}$~\cite{Wiringa1995_PRC51-38} and the three-nucleon (3N) interaction Urbana IX ~\cite{Pudliner1995_PRL74-4396}. It was found that commonly used Skyrme functionals overestimate the central density of these drops and the spin-orbit splitting of drops with 7-neutron~\cite{Pudliner1996_PRL76-2416}.
In Ref.~\cite{Smerzi1997_PRC56-2549} the ground state energy was studied for $N = 6$ neutron drops and the neutron pairing energy was discussed by comparison with Ref.~\cite{Pudliner1996_PRL76-2416}.
Later, more systematic studies have been performed for larger $N$ values with different external fields and different interactions using quantum Monte Carlo methods~\cite{Pederiva2004_NPA742-255,Gandolfi2011_PRL106-012501,Maris2013_PRC87-054318,Tews2016_PRC93-024305}.
Studies with the modern high precision chiral 2N interaction N$^3$LO \cite{Entem2003_PRC68-041001} and the 3N interaction N$^2$LO~\cite{Epelbaum2002_PRC66-064001} have been benchmarked with different \emph{ab initio} methods, including the no-core shell model~\cite{Barrett2013_PPNP69-131} and the coupled-cluster theory~\cite{Hagen2014_RPP77-096302}, and it was found that the results are consistent with each other~\cite{Potter2014_PLB739-445}.
However, by comparing these \emph{ab initio} calculations, one found a significant dependence on the selected interactions, especially on the 3N interactions~\cite{Maris2013_PRC87-054318,Potter2014_PLB739-445,Tews2016_PRC93-024305}.

On the other hand, various nonrelativistic and relativistic density functionals have been used to study neutron drops, and a strong linear correlation between the rms radii of neutron drops and the neutron skin thickness of $^{208}$Pb and $^{48}$Ca has been pointed out in Ref.~\cite{ZHAO-PW2016_PRC94-041302}.
Because of the uncertainty in the isovector part, there exists a large uncertainty in the results of neutron drops for the different functionals.

Recently, the self-consistent relativistic Brueckner-Hartree-Fock (RBHF) theory for finite nuclei has been established, and the results are in much better agreement with experimental data than the nonrelativistic calculations with the 2N interaction only \cite{SHEN-SH2016_ChPL33-102103,Shen2017}.
Indeed, it is known since more than 30 years that relativistic Brueckner-Hartree-Fock theory gives a much better description of the nuclear matter saturation properties than nonrelativistic BHF theories~\cite{Anastasio1983_PR100-327,Brockmann1984_PLB149-283,terHaar1987_PR149-207}.
In nonrelativistic many-body investigations on the influence of various types of 3N-interactions, it was found that a relativistic effect, the so-called $Z$-diagram, plays a major role~\cite{Zuo2002_NPA706-418}.

Having these progresses in mind, it is important to study the neutron drops in more detail in the framework of RBHF theory and compare the results with other nonrelativistic \emph{ab initio} calculations using various 2N or 2N + 3N interactions, as well as calculations using various density functionals.
This can also provide valuable insight to improve current relativistic density functionals.
In Ref.~\cite{Shen2018}, a systematic and specific pattern due to the tensor forces in the evolution of spin-orbit splittings based on RBHF theory is reported.

In this work, we investigate neutron drops confined in an external harmonic oscillator potential using relativistic Brueckner-Hartree-Fock theory, and present the numerical details and calculated results in detail.
In Sec. \ref{sec:theory}, we give a brief outline of the RBHF framework for neutron drops. The numerical details are discussed in Sec. \ref{sec:nd}. Results and discussion for neutron drops with an even number of neutrons from $N = 4$ to $50$ will be presented in Sec. \ref{sec:res}. Finally, a summary and perspectives for future investigations will be given in Sec. \ref{sec:sum}.


\section{Theoretical Framework}\label{sec:theory}

In this Section, we will outline the theoretical framework of relativistic Brueckner-Hartree-Fock theory for neutron drops.
For a detailed description of RBHF theory for finite nuclei, we refer to Refs.~\cite{SHEN-SH2016_ChPL33-102103,Shen2017}.

We start with a relativistic one-boson-exchange $NN$ interaction which describes the $NN$ scattering data~\cite{Machleidt1989_ANP19-189}:
\begin{align}
\mathscr{L}_{NNpv} &= -\frac{f_{ps}}{m_{ps}} \bar{\psi}\gamma^5\gamma^\mu
\psi \partial_\mu \varphi^{(ps)},  \notag \\
\mathscr{L}_{NNs} &= g_s \bar{\psi}\psi\varphi^{(s)}, \\
\mathscr{L}_{NNv} &= -g_v \bar{\psi}\gamma^\mu\psi\varphi_\mu^{(v)} - \frac{%
f_v}{4M}\bar{\psi}\sigma^{\mu\nu}\psi \left( \partial_\mu\varphi_\nu^{( v)}
- \partial_\nu \varphi_\mu^{(v)} \right),  \notag
\end{align}
where $\psi$ denotes the nucleon field. The bosons to be exchanged are
characterized by the index $\alpha$ and
include the pseudoscalar mesons ($\eta,\pi$) with a pseudovector ($pv$)
coupling, the scalar ($s$) mesons ($\sigma,\delta$), and the vector ($v$) mesons
($\omega,\rho$). For each pair, e.g., ($\eta,\pi$), the first (second) meson has
isoscalar (isovector) character. For the isovector mesons, the
field operator $\varphi_\alpha$ is replaced by $\vec{\varphi}_\alpha\cdot\vec{\tau}$
with $\vec{\tau}$ being the usual Pauli matrices in isospace.

The Hamiltonian is obtained through the Legendre transformation.
Considering the stationary case, the Hamiltonian can be expressed in the second quantized form as:
\begin{equation}
H=\sum_{kk^{\prime }}\langle k|T|k^{\prime }\rangle b_{k}^{\dagger
}b_{k^{\prime }}^{{}}+\frac{1}{2}\sum_{klk^{\prime }l^{\prime }}\langle
kl|V|k^{\prime }l^{\prime }\rangle b_{k}^{\dagger }b_{l}^{\dagger
}b_{l^{\prime }}^{{}}b_{k^{\prime }}^{{}},  \label{eq:hami}
\end{equation}%
where the relativistic matrix elements are given by
\begin{align}
\langle k|T|k^{\prime }\rangle & =\int d^{3}r\,\bar{\psi}_{k}(\mathbf{r})\left( -i%
\bm{\gamma}\cdot \nabla +M\right) \psi _{k^{\prime }}(\mathbf{r}), \\
\langle kl|V_{\alpha }|k^{\prime }l^{\prime }\rangle & =\int
d^{3}r_{1}d^{3}r_{2}\,\bar{\psi}_{k}(\mathbf{r}_{1})\Gamma _{\alpha }^{(1)}\psi
_{k^{\prime }}(\mathbf{r}_{1})  \notag \\
&~~~~~~~~\times D_{\alpha }(\mathbf{r}_{1},\mathbf{r}_{2})\bar{\psi}_{l}(\mathbf{r}_{2})\Gamma
_{\alpha }^{(2)}\psi _{l^{\prime }}(\mathbf{r}_{2}).
\end{align}
The indices $k,l$ run over an arbitrary complete basis of Dirac spinors with positive and negative energies, as, for instance, over the eigensolutions of a Dirac equation with potentials of Woods-Saxon shapes discussed in Refs.~\cite{ZHOU-SG2003_PRC68-034323,Meng2006,Shen2017}.
The two-body interaction $V$ contains contributions from the different mesons $\alpha$.
The interaction vertices for particles 1 and 2 are $\Gamma _{\alpha }^{(1)}$ and  $\Gamma _{\alpha }^{(2)}$:
\begin{subequations}\label{eq:gamma12}
\begin{align}
\Gamma _{s}=&~g_{s}, \\
\Gamma _{pv}=&~\frac{f_{ps}}{m_{ps}}\gamma ^{5}\gamma ^{i}\partial _{i}, \\
\Gamma _{v}^{\mu }=&~g_{v}\gamma ^{\mu }+\frac{f_{v}}{2M}\sigma ^{i\mu
}\partial _{i}.
\end{align}%
\end{subequations}
In the Bonn interaction, there is a form factor of monopole-type attached to each vertex.
It has the form in momentum space:
\begin{equation}
\frac{\Lambda _{\alpha }^{2}-m_{\alpha }^{2}}{\Lambda _{\alpha }^{2}+\mathbf{q}%
^{2}},  \label{eq:form}
\end{equation}%
where $\Lambda _{\alpha }$ is the cut-off parameter for meson $\alpha $ and $\mathbf{q}$ is the momentum transfer following Ref.~\cite{Machleidt1989_ANP19-189}.

The meson propagators $D_{\alpha }(\mathbf{r}_1,\mathbf{r}_2)$ are the retarded solutions of the Klein-Gordon equations in Minkowsky space.
For the Bonn interaction, this retardation effect was deemed to
be small and was ignored from the beginning~\cite{Machleidt1989_ANP19-189}. In this way, the $q_0$ integration can be carried out and we are left with the meson propagators in $r$-space:
\begin{eqnarray}
D_{\alpha }(\mathbf{r}_1,\mathbf{r}_2)&=&
\pm \int \frac{d^{3}q}{(2\pi )^{3}}\frac{1}{m_{\alpha }^{2}+\mathbf{q}^{2}}e^{i\mathbf{q}\cdot(\mathbf{r}_1-\mathbf{r}_2)}.
\label{eq:propa}
\end{eqnarray}%
The sign $-$ holds for scalar (and pseudoscalar) mesons and the sign $+$ for the vector fields.
Note that with the form factor in Eq.~(\ref{eq:form}), the meson propagators are no longer simple Yukawa functions, but they can be evaluated in analytic form~\cite{PhD_Serra2001}.

The matrix elements of the bare nucleon-nucleon interaction are very large and difficult to be used directly in nuclear many-body theory. Within Brueckner theory, the bare interaction is replaced by an effective interaction in the nuclear medium, the $G$-matrix. It takes into account short-range correlations by summing up all the ladder diagrams of the bare interaction~\cite{Brueckner1954_PR95-217,Brueckner1954_PR96-508} and it is deduced from the Bethe-Goldstone equation~\cite{Bethe1957_PRSA238-551},
\begin{widetext}
\begin{equation}
\langle ab|\bar{G}(W)|a^{\prime }b^{\prime }\rangle =\langle ab|\bar{V}%
|a^{\prime }b^{\prime }\rangle +\frac{1}{2}\sum_{cd}\langle ab|\bar{V}%
|cd\rangle \frac{Q(c,d)}{W-\varepsilon _{c}-\varepsilon _{d}}\langle cd|\bar{%
G}(W)|a^{\prime }b^{\prime }\rangle,
\label{eq:BG}
\end{equation}%
\end{widetext}
where $|a\rangle,|b\rangle$ are solutions of the relativistic Hartree-Fock equations,
$\langle ab|\bar{V}|a^{\prime }b^{\prime }\rangle=\langle ab|V|a^{\prime }b^{\prime }-b^{\prime }a^{\prime }\rangle$
are the antisymmetrized two-body matrix elements, $W$ is the starting
energy, and $\varepsilon _{c}$, $\varepsilon _{d}$ are the single-particle
energies of the two particles in the intermediate states. The Pauli operator
$Q(c,d)$ allows the scattering only to states $c$ and $d$ above the Fermi
surface. We also do not allow the scattering to states in the Dirac sea. Therefore $Q(c,d)$ is defined as
\begin{equation}
Q(c,d)=\begin{cases} 1, & \rm{for}~~\varepsilon_{c}>\varepsilon_{F}~and~
\varepsilon_{d}>\varepsilon_{F}, \\ 0, & \rm{otherwise}. \end{cases}
\label{eq:Q}
\end{equation}%

The single-particle motion fulfills the relativistic Hartree-Fock (RHF) equation, which in an external field reads
\begin{equation}
(T+U+U_{\rm ex})|a\rangle =e_{a}|a\rangle ,
\label{eq:rhf}
\end{equation}%
where $e_{a}=\varepsilon _{a}+M$ is the single-particle energy with the rest
mass of the nucleon $M$, and $U_{\rm ex}$ is the external field to confine the neutron drop.
The self-consistent single-particle potential $U$ is defined with the $G$-matrix by \cite{Baranger1969_Varenna40,Davies1969_PRC177-1519}:
\begin{equation}
U_{ab}=\frac{1}{2}\sum_{c=1}^{N}\langle ac|\bar{G}(\varepsilon
_{a}+\varepsilon _{c})+\bar{G}(\varepsilon _{b}+\varepsilon _{c})|bc\rangle ,
\label{eq:Uhh}
\end{equation}%
if $|a\rangle $ and $|b\rangle $ are both hole (i.e. occupied) states, and
\begin{equation}
U_{ab}=\sum_{c=1}^{N}\langle ac|\bar{G}(\varepsilon _{a}+\varepsilon
_{c})|bc\rangle ,
\label{eq:Uph}
\end{equation}%
if $|a\rangle $ is a hole state and $|b\rangle $ is a particle (i.e. unoccupied) state, and
\begin{equation}
U_{ab}=\frac{1}{2}\sum_{c=1}^{N}\langle ac|\bar{G}(\varepsilon _{a}^{\prime
}+\varepsilon _{c})+\bar{G}(\varepsilon _{b}^{\prime }+\varepsilon
_{c})|bc\rangle ,
\label{eq:Upp}
\end{equation}%
if $|a\rangle $ and $|b\rangle $ are both particle states.
In the above expression, the summation index $c$ goes through $N$-neutron occupied states.

In the above expressions, $\varepsilon$ labels the self-consistent single-particle energies, while $\varepsilon^{\prime }$ is somewhat uncertain \cite{Davies1969_PRC177-1519}.
The matrix elements of the self-consistent potential $U_{ab}$ with both states $|a\rangle$ and $|b\rangle$ above the Fermi level are not well defined in the Brueckner-Hartree-Fock theory.
Different choices have been proposed in the literature \cite{Rajaraman1967_RMP39-745,Davies1969_PRC177-1519}.
Following the discussions in Ref.~\cite{Davies1969_PRC177-1519,Shen2017}, we choose $\varepsilon^{\prime }_a=\varepsilon^{\prime }_b=\varepsilon_{1s1/2}$ fixed as the lowest energy of the occupied states in the Fermi sea.

\section{Numerical details}\label{sec:nd}

We use the Bonn interaction which has been adjusted to the $NN$ scattering data in Ref.~\cite{Machleidt1989_ANP19-189}.
The neutron drops will be confined in a spherical harmonic oscillator (HO) trap, i.e., the external field in Eq.~(\ref{eq:rhf}) is
\begin{equation}
U_{\rm ex} = \frac{1}{2} M\omega^2 r^2
\label{eq:uex}
\end{equation}%
where the strength is chosen as $\hbar\omega = 10$ MeV if without specification. In contrast to the relativistic Brueckner-Hartree-Fock calculations for self-bound nuclei in Refs.~\cite{SHEN-SH2016_ChPL33-102103,Shen2017}, where we had to introduce a center of mass correction, this is not necessary here, because in the external field translational symmetry is lost.
The initial basis is the Dirac Woods-Saxon (DWS) basis \cite{ZHOU-SG2003_PRC68-034323}, and during the RBHF iteration it will be transformed to the self-consistent RHF basis as explained in Ref.~\cite{Shen2017}.
The DWS basis is obtained by solving the spherical Dirac equation in a box with the box size $R_{\rm box} = 8$ fm and a mesh size $dr=0.05$~fm.
The way to solve the BG equation~(\ref{eq:BG}) is the same as in Refs.~\cite{SHEN-SH2016_ChPL33-102103,Shen2017}, except that now only the isospin channel $T_z = 1$ is included.

It is well known that the bare $NN$ interaction contains a repulsive core and a strong tensor part connecting the nucleons below the Fermi surface to the states with high momentum in the continuum.
In order to take this coupling fully into account, one needs a relatively large basis space.
The convergence in finite nuclei has been confirmed in Refs.~\cite{SHEN-SH2016_ChPL33-102103,Shen2017}, in which reasonable convergence is achieved near an energy cut-off $\varepsilon_{\rm cut} = 1.1$ GeV.
For the neutron drops, we will carry out the same check.

\begin{figure}
\includegraphics[width=8cm]{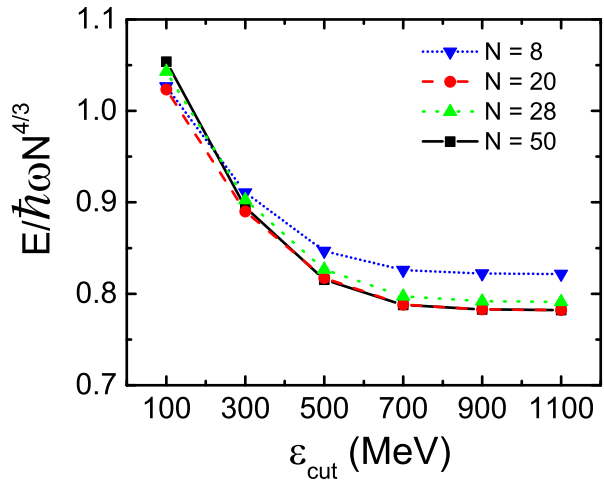}
\caption{(Color online) Total energy (in units of $\hbar\omega N^{4/3}$) of neutron drops with $N = 8,\,20,\,28$ and $50$ in a HO trap ($\hbar\omega = 10$ MeV) as a function of the energy cut-off $\varepsilon_{\rm cut}$ calculated by RBHF theory using the interaction Bonn A.}
\label{fig1}
\end{figure}

Fig.~\ref{fig1} shows the total energy divided by $\hbar\omega N^{4/3}$ of neutron drops with $N = 8,\,20,\,28$ and $50$ in a HO trap ($\hbar\omega = 10$ MeV) as a function of the energy cut-off $\varepsilon_{\rm cut}$ calculated by RBHF theory using the Bonn A interaction. The factor $\hbar\omega N^{4/3}$ is based on the consideration that in Thomas Fermi approximation~\cite{Ring1980} the total energy for a non-interacting $N$-Fermion system in a HO trap is given by
\begin{equation}\label{eq:Etf}
E = \frac{3^{4/3}}{4}\hbar\omega N^{4/3}\approx 1.082\,\hbar\omega N^{4/3}.
\end{equation}
In other words, all the energy below the line $E/\hbar\omega N^{4/3}\approx 1.082$ corresponds to binding induced through the nuclear force. This intrinsic binding energy grows linearly with $\hbar\omega$. With increasing neutron number of the drops we observe a saturation, but in contrast to the nuclear case where the binding energy grows roughly with the mass number $A$, here it grows for large $N\geq 20$ with $N^{4/3}$.

It can be seen from Fig.~\ref{fig1} that, first, the convergence with the energy cut-off does not depend on the particle numbers. Second, the convergence is achieved already at $\varepsilon_{\rm cut} = 900$ MeV, which is faster than $\varepsilon_{\rm cut} = 1100$ MeV in finite nuclei~\cite{SHEN-SH2016_ChPL33-102103,Shen2017}. This is because the tensor term plays a role in connecting the nucleons below the Fermi surface to the states with high momentum, and only the $T = 1$ term shows in neutron drops.

In order to compare the speed of convergence between neutron drops and finite nuclei, from the total energy of the system we define the following convergence rate
\begin{equation}\label{eq:mu}
\mu(\varepsilon_{\rm cut}) = \frac{E(\varepsilon_{\rm cut}-200~\text{MeV})-E(\varepsilon_{\rm cut})}{E(\varepsilon_{\rm cut})-E(\varepsilon_{\rm cut}+200~\text{MeV})},
\end{equation}
The larger the quantity $\mu$ is, the faster the convergence is.

\begin{figure}
\includegraphics[width=8cm]{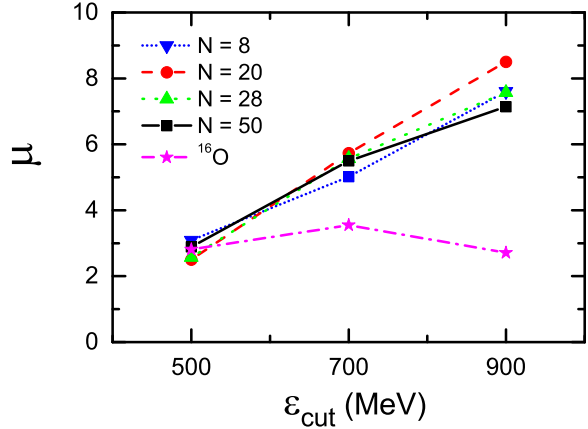}
\caption{(Color online) Convergence rate $\mu$ for neutron drops with $N = 8,\,20,\,28$ and $50$ in a HO trap ($\hbar\omega = 10$ MeV) and for the nucleus $^{16}$O as a function of the energy cut-off $\varepsilon_{\rm cut}$ calculated by RBHF theory using the Bonn A interaction.}
\label{fig2}
\end{figure}

In Fig.~\ref{fig2}, we show the convergence rate $\mu$ calculated with RBHF theory using the interaction Bonn A for neutron drops with $N = 8,\,20,\,28$ and $50$ in a HO trap with $\hbar\omega = 10$ MeV, and for the nucleus $^{16}$O (from Refs.~\cite{SHEN-SH2016_ChPL33-102103,Shen2017}).
It can be seen that at $\varepsilon_{\rm cut} = 500$ MeV, the convergence rates between different neutron drops and $^{16}$O are similar. As $\varepsilon_{\rm cut}$ increases, the convergence rates of neutron drops increase linearly, and they are similar for neutron drops with different neutron numbers. On the other hand, the convergence rate of $^{16}$O does not change too much as $\varepsilon_{\rm cut}$ increases and it is much slower than that of neutron drops.

\section{Results and discussion}\label{sec:res}

\subsection{Total energy}

\begin{table}[!th]
\caption{Total energy $E$ and rms radius $R_N$ of $N$-neutron drops in a HO trap ($\hbar\omega = 10$ MeV) calculated by RBHF theory using the interactions Bonn A, B, and C.}
\label{table1}
\centering
\begin{ruledtabular}
\begin{tabular}{crrrrrr}
& \multicolumn{2}{c}{Bonn A} & \multicolumn{2}{c}{Bonn B} & \multicolumn{2}{c}{Bonn C} \\
$N$ & E (MeV) & $R_N$ (fm) & E (MeV) & $R_N$ (fm) & E (MeV) & $R_N$ (fm) \\
\hline
4  &  62.6 & 2.51 &  62.6 & 2.51 &  62.7 & 2.51 \\
6  &  94.2 & 2.51 &  94.3 & 2.51 &  94.4 & 2.51 \\
8  & 130.0 & 2.61 & 130.2 & 2.61 & 130.3 & 2.61 \\
10 & 183.5 & 2.73 & 183.8 & 2.74 & 183.9 & 2.74 \\
12 & 231.2 & 2.80 & 231.6 & 2.81 & 231.8 & 2.81 \\
14 & 275.4 & 2.84 & 275.9 & 2.85 & 276.2 & 2.85 \\
16 & 320.2 & 2.89 & 321.0 & 2.90 & 321.4 & 2.90 \\
18 & 373.2 & 2.97 & 374.3 & 2.98 & 374.7 & 2.98 \\
20 & 418.1 & 3.02 & 419.3 & 3.03 & 419.7 & 3.03 \\
22 & 485.5 & 3.08 & 487.0 & 3.08 & 487.4 & 3.08 \\
24 & 546.9 & 3.12 & 548.7 & 3.13 & 549.2 & 3.13 \\
26 & 606.4 & 3.16 & 608.5 & 3.17 & 609.1 & 3.17 \\
28 & 663.9 & 3.19 & 666.3 & 3.20 & 666.9 & 3.20 \\
\end{tabular}
\end{ruledtabular}
\end{table}

In Table~\ref{table1} we list the total energies and root-mean-square (rms) radii of $N$-neutron drops in a HO trap ($\hbar\omega = 10$ MeV) calculated by RBHF theory using the interactions Bonn A, B, and C~\cite{Machleidt1989_ANP19-189}.
The results given for the interactions Bonn A, B and C are very similar. This can be understood by the fact that the main difference among the three Bonn interactions is the strength of the $T = 0$ tensor force~\cite{Machleidt1989_ANP19-189}, which has no influence on the neutron-neutron states with $T = 1$. This result is also in consistent with the finding in pure neutron matter, where the equation of state calculated by RBHF with Bonn A, B and C interactions are very close \cite{Li-GQ1992_PRC45-2782}.

\begin{figure}[!thbp]
\includegraphics[width=8cm]{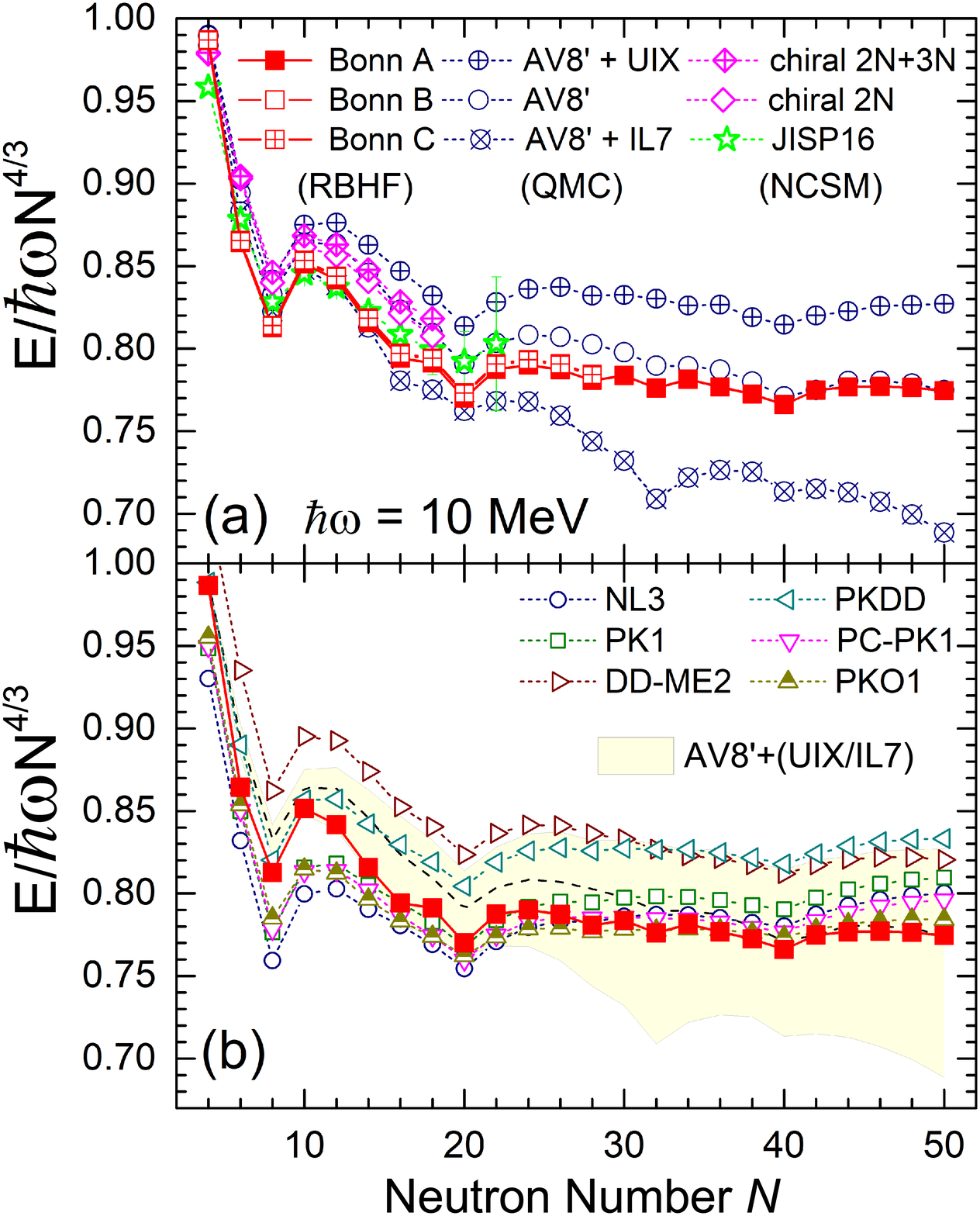}
\caption{Total energy in units of $\hbar\omega N^{4/3}$ for $N$-neutron drops in a HO trap ($\hbar\omega = 10$ MeV) calculated by RBHF theory using the interaction Bonn A: (a) in comparison with QMC calculations~\cite{Gandolfi2011_PRL106-012501,Maris2013_PRC87-054318} using the interactions AV8' + UIX, AV8', and AV8' + IL7, with NCSM calculations~\cite{Potter2014_PLB739-445,Maris2013_PRC87-054318} using chiral 2N + 3N forces, chiral 2N forces, and the interaction JISP16. (b) in comparison with results based on relativistic density functionals \cite{ZHAO-PW2016_PRC94-041302,Long2006_PLB640-150}. The shaded area indicates the QMC results.
}
\label{fig3}
\end{figure}

In Fig.~\ref{fig3}, we show the total energy in units of $\hbar\omega N^{4/3}$ for $N$-neutron drops (with $N$ from 4 to 50) in a HO trap ($\hbar\omega = 10$ MeV) calculated by RBHF theory using the Bonn interactions.
For the cases of  open shells, the filling approximation is used.
The results are compared with quantum Monte-Carlo (QMC) calculations~\cite{Gandolfi2011_PRL106-012501,Maris2013_PRC87-054318}  based on the interactions AV8' + UIX, AV8', and AV8' + IL7, with no-core shell model (NCSM) calculations~\cite{Potter2014_PLB739-445,Maris2013_PRC87-054318} based on chiral 2N + 3N forces, on chiral 2N force, and on JISP16, with calculations using relativistic density functionals~\cite{ZHAO-PW2016_PRC94-041302,Long2006_PLB640-150}.

As has already been discussed above, the results of Bonn A, B, and C are very similar. Therefore, in later discussions we will use Bonn A only.
By comparing with QMC and NCSM calculations, the results of RBHF with the interaction Bonn A are similar to the results of the JISP16 interaction, and AV8' + IL7 (for $N\leq 14$), and getting closer to AV8' for $N\geq 20$.
This similarity is favourable as JISP16 is a phenomenological nonlocal $NN$ interaction which can reproduce scattering data and describe well for light nuclei \cite{Shirokov2007_PLB644-33,Maris2009_PRC79-014308}.
On the other hand, AV8' + IL7 gives better description for light nuclei up to A = 12 than AV8' or AV8' + UIX, but gives too much over-binding for pure neutron matter at higher densities \cite{Sarsa2003_PRC68-024308,Maris2013_PRC87-054318}.

\begin{figure}[!thbp]
\includegraphics[width=8cm]{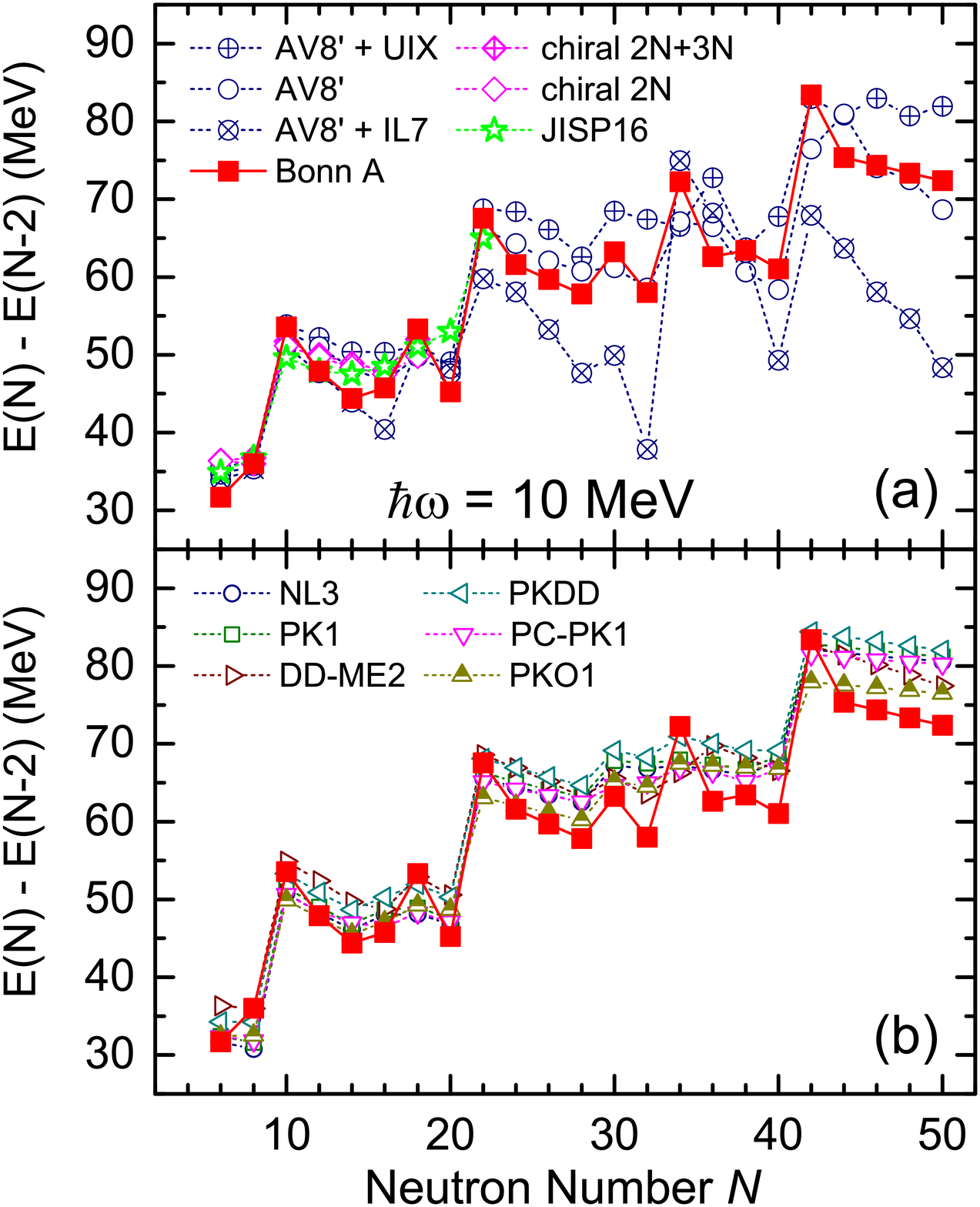}
\caption{Two neutron energy difference of $N$-neutron drops in a HO trap ($\hbar\omega = 10$ MeV) calculated by RBHF theory using Bonn A interaction. (a) In comparison with QMC calculations~\cite{Gandolfi2011_PRL106-012501,Maris2013_PRC87-054318} using the interactions AV8' + UIX, AV8', and AV8' + IL7, with NCSM calculations using the interactions chiral 2N + 3N force, chiral 2N force~\cite{Potter2014_PLB739-445}, and JISP16~\cite{Maris2013_PRC87-054318}. (b) In comparison with relativistic density functionals \cite{ZHAO-PW2016_PRC94-041302}.}
\label{fig4}
\end{figure}

In comparison with relativistic density functional calculations, we took four types of functionals, which cover a wide range of relativistic density functionals presently on the market:
\begin{enumerate}
  \item non-linear meson couplings: NL3~\cite{Lalazissis1997}, PK1~\cite{Long2004_PRC69-034319};
  \item density-dependent meson couplings: DD-ME2~\cite{Lalazissis2005_PRC71-024312}, PKDD~\cite{Long2004_PRC69-034319};
  \item point-couplings: PC-PK1~\cite{Zhao2010};
  \item functional for RHF-calculations: PKO1~\cite{Long2006_PLB640-150} (which includes tensor force).
\end{enumerate}
As there is no pairing in the RBHF calculation, we do not include pairing neither in the relativistic density functional calculations.
We would like to mention that generally by including pairing effects, the binding energies of open shell neutron drops would get larger, while for closed shell drops they are unchanged.
The overall energy as a function of neutron number N will be smoother as demonstrated in Ref.~\cite{ZHAO-PW2016_PRC94-041302}.
However, the effect is too small to be significant in the figure, therefore we will not plot it out and more importantly, for the purpose of consistency with RBHF calculation, we will use the results without pairing in the rest discussions.

From Fig. \ref{fig3}(b) it can be seen that the binding energies given by RBHF are generally bigger than those given by DD-ME2 and PKDD.
For $N = 8$, RBHF is close to PKDD, but getting closer to PK1 from $N = 14$ to 26, and closer to PC-PK1, NL3, and PKO1 from $N = 28$ to 36. From $N = 20$ on, the results of RBHF and DD-ME2 are close to a horizontal line, while the others have a small tendency of increasing. The microscopic results obtained by RBHF can be a guidance for future density functionals.
For example, the neutron-neutron interaction might be too repulsive in DD-ME2, whereas it might be too attractive in NL3 when the neutron number $N$ is small and then become repulsive as $N$ becomes large.

Since in these calculations the $\hbar\omega = 10$ MeV HO external field is chosen, they all show the HO magic number 8, 20, and 40.
Beside the above magic numbers, the results of RBHF indicate a sub-shell closure at $N = 32$, similar as the results of AV8' + IL7.
The sub-shell closure at $N = 32$ is not significant for AV8', and does not exist for AV8' + UIX.
For the $N = 28$ sub-shell closure, results of Bonn A and AV8' + UIX show a small hint, while AV8' and AV8' + IL7 do not show it.
On the other hand, all the relativistic density functionals only show the HO magic number 8, 20, 40, and no clear sub-shells closures for $N = 28$ or 32.

In order to see the shell structure more clearly, we present in Fig.~\ref{fig4} the negative two neutron separation energies $E(N) - E(N-2)$ for the above calculations. The HO magic number 8, 20, 40 are clearly shown in all calculations.
But the traditional sub-shell at $N = 28$ in a finite system does not show up evidently in neutron drops.
On the other hand, the results of AV8' + IL7 shows a prominent sub-shell closure at $N = 32$, while RBHF with Bonn A shows a modest but also clear closure at that neutron number.

By looking into Fig.~\ref{fig4} (b), it can be seen that the results of relativistic density functionals are much smoother than those of the \emph{ab initio} calculations.
In particular it is interesting to see that these density functionals do not show clear sub-shell structure at $N = 32$ and only a small closure at $N = 28$.
The sub-shell closure is related to the underlying single-particle spectra.
Taking the $N = 32$ sub-shell as an example, the $2p_{3/2}$ state is just fully occupied and from $N = 34$ on the $1f_{5/2}$ state (for certain cases $2p_{1/2}$) will begin to be occupied.
Therefore, the gap between single-particle states $1f_{5/2}$ (or $2p_{1/2}$) and $2p_{3/2}$ is a reflection of how strong the $N = 32$ sub-shell is, see also the discussions on the single-particle energies in Subsection \ref{sec:spe}.
For RBHF with Bonn A, this gap is $3.047$ MeV in the $N = 34$ drop, while other relativistic density functionals give values ranging from $0.401$ MeV (NL3) to $2.127$ MeV (DD-ME2), which are much smaller than that of RBHF.
Therefore the $N = 32$ sub-shell closure given by RBHF is stronger than those of relativistic density functionals.
This might be a hint that some parts of the effective Lagrangian are missing in these models.
However, in order to understand the underlying detail, a decomposition of the $G$-matrix into different channels (scalar, vector, tensor, and so on) and a careful comparison with various density functionals are indispensable. Work along this direction is in progress.

\subsection{Radii}

\begin{figure}[!t]
\includegraphics[width=8cm]{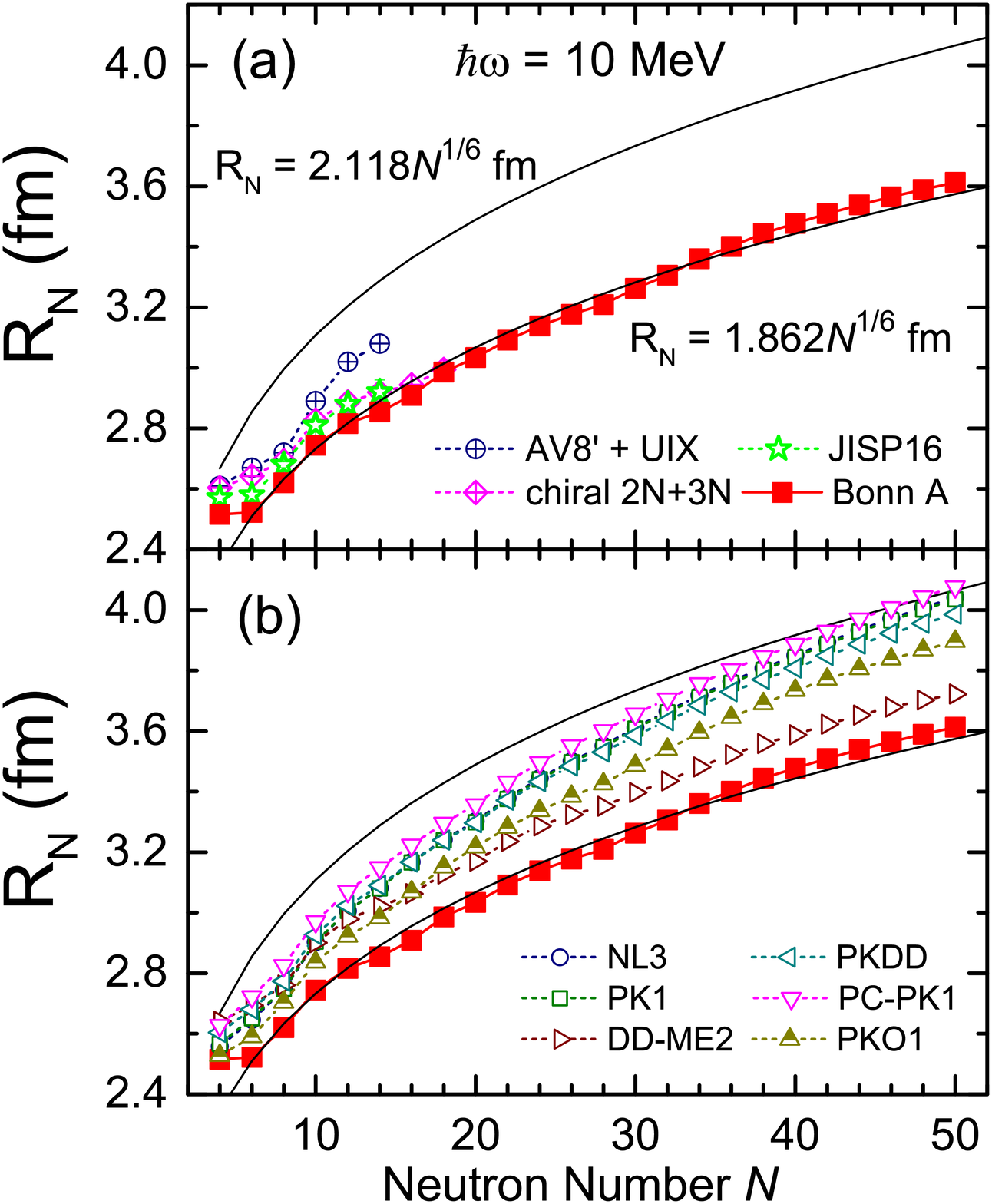}
\caption{Radii of N-neutron drops in a HO trap ($\hbar\omega = 10$ MeV) calculated by RBHF theory using the interaction Bonn A: (a) in comparison with QMC calculations using the interaction AV8' + UIX~\cite{Gandolfi2011_PRL106-012501}, with NCSM calculations~\cite{Potter2014_PLB739-445,Maris2013_PRC87-054318} using the chiral 2N + 3N force and the interaction JISP16; (b) in comparison with relativistic density functionals \cite{ZHAO-PW2016_PRC94-041302}. Further details are given in the text.}
\label{fig5}
\end{figure}

Fig.~\ref{fig5} shows the rms radii of $N$-neutron drops in a HO trap ($\hbar\omega = 10$ MeV) calculated in the framework of RBHF theory using the interaction Bonn A.
In the upper panel the results are compared with QMC calculations based on the interactions AV8' + UIX~\cite{Gandolfi2011_PRL106-012501}, with NCSM calculations~\cite{Potter2014_PLB739-445,Maris2013_PRC87-054318} based on the chiral 2N + 3N force and based on the JISP16 force.
In the lower panel these results are compared with calculations using relativistic density functionals.
The black line in the upper and the lower panels $R_N = 2.118 N^{1/6}$ fm is obtained by solving for free Fermions in a $\hbar\omega = 10$ MeV HO trap using the Thomas-Fermi approximation, which can be derived as
\begin{equation}
R_N = \left(\frac{3^{4/3}}{4}\frac{\hbar}{M\omega}\right)^{1/2} N^{1/6}.
\end{equation}
For $M = 938.926$ MeV  and $\hbar\omega = 10$ MeV, one finds a factor $2.118$ fm in front of $N^{1/6}$.
The black line $R_N = 1.862 N^{1/6}$ fm is obtained by fitting to the results of Bonn A from $N$ = 6 to 50.

Generally, all the calculated radii fulfill the relationship $N^{1/6}$ as a function of $N$.
In all the selected calculations, RBHF with Bonn A gives the smallest radii.
By comparing with other calculations in Fig.~\ref{fig5}(a) and Fig.~\ref{fig3}(a), we find that while AV8' + UIX gives the smallest binding energies, it also gives the largest radii. Even though the energies given by JISP16 are similar to those of Bonn A, the radii given by JISP16 are larger than those of Bonn A. The radii of relativistic density functionals in Fig.~\ref{fig5}(b) are much larger than those of RBHF, even though some of their binding energies are larger than RBHF before $N = 20$ in Fig.~\ref{fig3}(b).

It is known (Ref.~\cite{Niksic2002}) that the relativistic density functionals without density-dependence in the isovector channel show too large neutron radii in realistic nuclei and we observe this for the neutron drops too.
The relation between the slope parameter $L$ and the neutron skin is well known.
For neutron drops this is also discussed in Ref.~\cite{ZHAO-PW2016_PRC94-041302}.

\begin{table}[!th]
\caption{Rms radius $R_N$ of $N = 50$ neutron drop in a HO trap ($\hbar\omega = 10$ MeV) calculated by RBHF theory using the interaction Bonn A. The asymmetry energy $a_{\rm sym}$ and slope parameter $L$ calculated in nuclear matter \cite{Alonso2003,VanDalen2004,Katayama2013} have also been listed. They are compared with results of relativistic density functionals NL3~\cite{Lalazissis1997}, PK1~\cite{Long2004_PRC69-034319}, DD-ME2~\cite{Lalazissis2005_PRC71-024312}, PKDD~\cite{Long2004_PRC69-034319}, PC-PK1~\cite{Zhao2010}, and PKO1~\cite{Long2006_PLB640-150}.}
\label{table2}
\centering
\begin{ruledtabular}
\begin{tabular}{crrr}
 & $R_{N=50}$ (fm) & $a_{\rm sym}$ (MeV) & $L$ (MeV)  \\
\hline
Bonn A & $3.61$ & $34.8$ \cite{Alonso2003,VanDalen2004,Katayama2013} & $71$ \cite{Katayama2013} \\
NL3 & $4.04$ & $36.6$ & $119$ \\
PK1 & $4.04$ & $37.6$ & $116$ \\
DD-ME2 & $3.72$ & $32.3$ & $51$ \\
PKDD & $3.99$ & $36.8$ & $90$ \\
PC-PK1 & $4.08$ & $35.6$ & $113$ \\
PKO1 & $3.90$ & $34.4$ & $98$ \\
\end{tabular}
\end{ruledtabular}
\end{table}

For a better comparison, we list the radius of the $N = 50$ neutron drop calculated by RBHF theory using the interaction Bonn A in table~\ref{table2}.
The asymmetry energy $a_{\rm sym}$ and slope parameter $L$ calculated in nuclear matter \cite{Alonso2003,VanDalen2004,Katayama2013} have also been listed.
They are compared with results of the relativistic density functionals. It can be seen that in general, the radius of a neutron drop is large if $a_{\rm sym}$ or $L$ is large, although in detail small discrepancies exist.
For example, DD-ME2 gives the smallest $a_{\rm sym}$ and $L$, and its radius is indeed the smallest among those of relativistic density functionals, but still larger than that of Bonn A.
The radius of PC-PK1 is the largest, and its $a_{\rm sym}$ or $L$ is large, but not the largest, which is slightly smaller than those of NL3 and PK1.

\begin{figure}[!thbp]
\includegraphics[width=8cm]{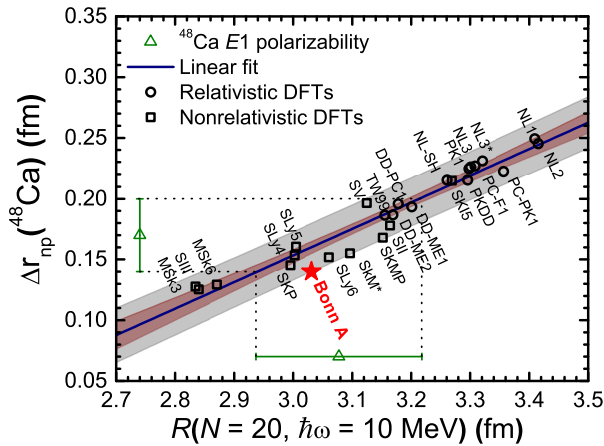}
\caption{Neutron skin thickness $\Delta r_{np}$ of $^{48}$Ca and the rms radius $R$ of $N = 20$ neutron drop in a HO trap ($\hbar\omega = 10$ MeV) calculated by RBHF theory using the interaction Bonn A (red star), in comparison with results obtained by various density functionals~\cite{ZHAO-PW2016_PRC94-041302}.
The datum of $\Delta r_{np}$ is obtained by measuring the electric dipole polarizability of $^{48}$Ca~\cite{Birkhan2017}.
The blue line is the linear fit to the results of density functionals, and the inner (outer) colored
regions depict the 95\% confidence (prediction) intervals of the linear regression~\cite{ZHAO-PW2016_PRC94-041302}.}
\label{fig6}
\end{figure}

In Ref.~\cite{ZHAO-PW2016_PRC94-041302}, a strong linear correlation has been found between the neutron skin thickness $\Delta r_{np}$ and the rms radius $R_N$ of $N$-neutron drops in an external HO field.
Fig.~\ref{fig6} shows the linear correlation between the neutron skin thickness of $^{48}$Ca and the radius of $N = 20$ neutron drops in a $\hbar\omega = 10$ MeV HO external field as given in Ref.~\cite{ZHAO-PW2016_PRC94-041302}.
The black circle and square symbols are calculated with different nonrelativistic and relativistic density functionals, and the blue line is obtained by fitting to these results \cite{ZHAO-PW2016_PRC94-041302}.
The inner (outer) colored regions depict the 95\% confidence (prediction) intervals of the linear regression.

The red star in Fig.~\ref{fig6} is calculated by RBHF theory using the interaction Bonn A.
The datum of the neutron skin thickness of $^{48}$Ca is obtained by measuring the electric dipole polarizability in Ref.~\cite{Birkhan2017}.
It can be seen that the neutron skin thickness of $^{48}$Ca given by RBHF $\Delta r_{np} = 0.14$ fm is located within the error bar of experimental data, which is also consistent with the $0.12 \leq \Delta r_{np} \leq 0.15$ fm given by coupled-cluster calculations using the interaction NNLO$_{\rm sat}$ \cite{Hagen2015_NatP3529}.

\begin{figure}[!thbp]
\includegraphics[width=8cm]{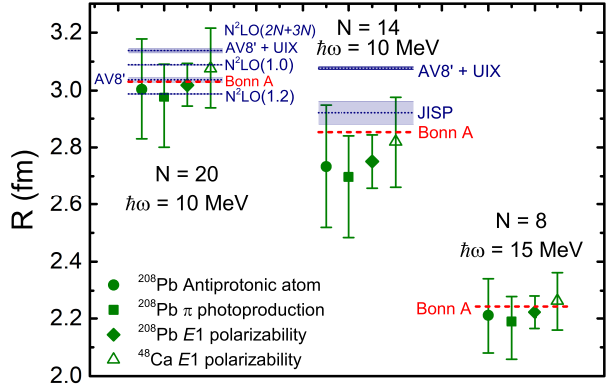}
\caption{Radii for $N = 20, 14$, and $8$ neutron drops calculated by RBHF theory using the interaction Bonn A (red dash), in comparison with data (green symbols) determined from the linear correlations with the neutron skin thicknesses of $^{208}$Pb and $^{48}$Ca~\cite{ZHAO-PW2016_PRC94-041302}, and other \emph{ab initio} calculations (blue dot)~\cite{Gandolfi2011_PRL106-012501,Maris2013_PRC87-054318,Tews2016_PRC93-024305}.
Blue colored regions denote theoretical uncertainties.}
\label{fig7}
\end{figure}

Apart from the linear correlation between $\Delta r_{np}$ of $^{48}$Ca and radius of $N = 20$ neutron drops in Fig.~\ref{fig6}, similar correlations can be found in other cases, for example for $\Delta r_{np}$ of $^{208}$Pb or other numbers of $N$.
Using these linear correlations, the experimental data of neutron skins of $^{48}$Ca and $^{208}$Pb can be mapped to the data of radii of neutron drops with different numbers of $N$ \cite{ZHAO-PW2016_PRC94-041302}, and the results are shown with green symbols in Fig.~\ref{fig7}.
In this way, the study on the neutron skin of heavy nuclei can be linked to the study of the radius of neutron drops, while the latter is much easier to be accessed by different \emph{ab initio} calculations.

In Fig.~\ref{fig7}, we show the radii for $N = 20, 14$, and $8$ neutron drops calculated by RBHF theory using the interaction Bonn A (red dashed lines), in comparison with data (green symbols) determined from the linear correlations with the neutron skin thicknesses of $^{208}$Pb and $^{48}$Ca~\cite{ZHAO-PW2016_PRC94-041302}, and other \emph{ab initio} calculations (blue dotted lines)~\cite{Gandolfi2011_PRL106-012501,Maris2013_PRC87-054318,Tews2016_PRC93-024305}.
For $\Delta r_{np}$ of $^{208}$Pb, the data come from different measurements with antiprotonic atoms \cite{Klos2007_PRC76-014311} (circle), pion photoproduction \cite{Tarbert2014_PRL112-242502} (square), and electric dipole polarizability \cite{RocaMaza2013_PRC88-024316} (diamond); for $\Delta r_{np}$ of $^{48}$Ca, the datum comes from the measurement of the electric dipole polarizability~\cite{Birkhan2017} (triangle).
For the local chiral forces N$^2$LO from Refs.~\cite{Gezerlis2013_PRL111-032501,Lynn2016_PRL116-062501}, we present the results of a two-body force with a cutoff $R_0 = 1.0$ and $1.2$ fm, and a two-body plus three-body force ($2N+3N$) with a cutoff $R_0 = 1.2$ fm \cite{Tews2016_PRC93-024305}. Theoretical uncertainties are denoted by blue colored regions.
There is no particular reason to choose $N = 20, 14$, and $8$ neutron drops, as long as the central density of the neutron drop does not differ too much from the saturation density ($\approx 0.16$ fm$^{-3}$) \cite{ZHAO-PW2016_PRC94-041302}.

It can be seen that the radii obtained in RBHF calculations with the interaction Bonn A are in good agreement with the data determined from the linear correlations with the neutron skin thicknesses. In comparison with other \emph{ab initio} calculations, AV8' + UIX shows more repulsion and gives larger radii, as expected from the energies shown in Fig.~\ref{fig3}.
For the $2N$ local chiral forces N$^2$LO, the softer interaction with a cut-off radius $R_0 = 1.2$ fm gives a smaller radius and the harder one with $R_0 = 1.0$ fm gives a larger radius. When including the $3N$ force for N$^2$LO, the radius gets larger by 0.05 fm and is in the same position as AV8' + UIX in Fig.~\ref{fig7}.

\subsection{Density distribution}

\begin{figure}[!thbp]
\includegraphics[width=8cm]{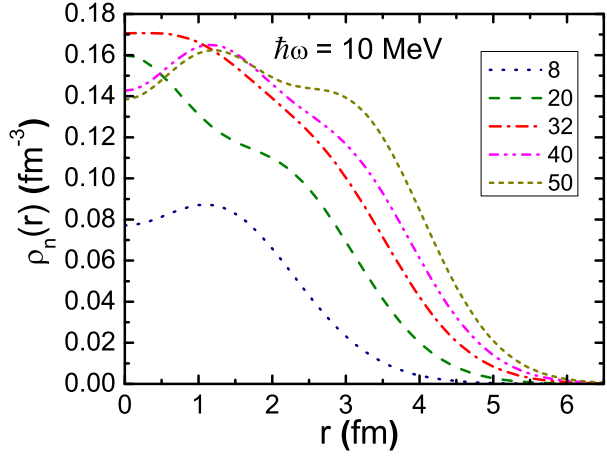}
\caption{Density distributions of $N$-neutron drops in a HO trap ($\hbar\omega = 10$ MeV) calculated by RBHF theory using the interaction Bonn A.}
\label{fig8}
\end{figure}

Fig.~\ref{fig8} shows the density distributions of $N$-neutron drops in a HO trap ($\hbar\omega = 10$ MeV) calculated by RBHF theory using the interaction Bonn A.
With given HO strength, the neutron density gets saturated around $0.14 - 0.17$ fm$^{-3}$.
For neutron drops with $N = 40$ or $N=50$ there is a bubble structure in the center.
This can be understood from the occupation of single-particle states.
Near $N = 20$, the $2s_{1/2}$ state has just been occupied and it has a large contribution to the central density.
From $N = 20$ to $N=50$, the $1f,\,2p,$ and $1g_{9/2}$ states start to be occupied and because their angular momentum $l\neq 0$, the density begins to shift outward.
Similar as the proton bubble structure in the $^{34}$Si, where the proton $2s_{1/2}$ state is empty and just to be occupied in the next nucleus $^{36}$S \cite{Mutschler2017}.

\subsection{Single-particle potential}

The single-particle potential in RHF equation (\ref{eq:rhf}) is a nonlocal potential.
However, for a given single-particle wave function, one can construct an equivalent local potential for this state by using the Dirac equation. For spherical symmetry, one has the radial equation,
\begin{equation}
\left(
\begin{array}{cc}
M+\Sigma(r) & -\frac{d}{dr}+\frac{\kappa}{r} \\
\frac{d}{dr}+\frac{\kappa}{r} & -M+\Delta(r)
\end{array}
\right) \left(
\begin{array}{c}
F_a(r) \\
G_a(r) \\
\end{array}
\right) =e_a \left(
\begin{array}{c}
F_a(r) \\
G_a(r) \\
\end{array}
\right),
\label{eq:direq}
\end{equation}
where $\Sigma=V+S$ and $\Delta=V-S$ are the sum and the difference of vector and
scalar potentials, the quantum number $\kappa$ is defined as $\kappa = \pm(j+1/2)$ for $j=l\mp1/2$.
Then one finds
\begin{align}
\Sigma_a(r) =&~e_a - M + \left(\frac{dG_a(r)}{dr}-\frac{\kappa}{r}G_a(r)\right)F^{-1}_a(r), \\
\Delta_a(r) =&~e_a + M - \left(\frac{dF_a(r)}{dr}+\frac{\kappa}{r}F_a(r)\right)G^{-1}_a(r).
\end{align}

\begin{figure}[!thbp]
\includegraphics[width=8cm]{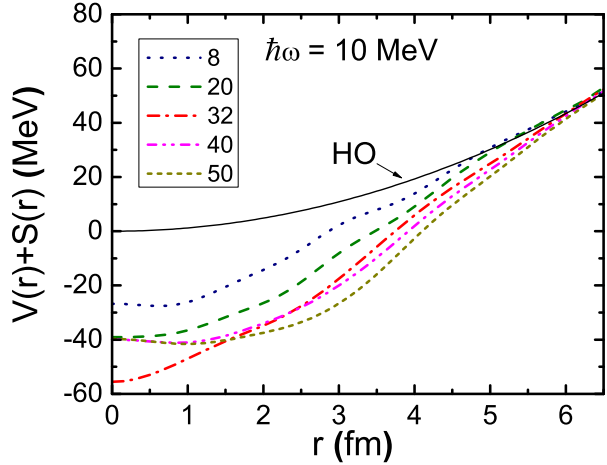}
\caption{Localized $1s_{1/2}$ single-particle potential of $N$-neutron drops in a HO trap ($\hbar\omega = 10$ MeV) calculated by RBHF theory using the interaction Bonn A.
The black line is the external HO potential.}
\label{fig9}
\end{figure}

Fig.~\ref{fig9} shows this localized  single-particle potential $\Sigma_{1s1/2}(r)$ for the $1s_{1/2}$ state of $N$-neutron drops in a HO trap ($\hbar\omega = 10$ MeV) calculated by RBHF theory using the interaction Bonn A.
As $r$ increases, the single-particle potentials approach to the external HO potential.
The central potential is negative and decreases as $N$ increases.
This is a consequence of the attractive interaction between the neutrons.
Similar to the density distribution shown in Fig.~\ref{fig8}, the single-particle potential saturates as $N$ increases to $20$ and the potential depth with respect to the potential of the HO trap is about $-40$ MeV.

\subsection{Single-particle energies}\label{sec:spe}

\begin{figure}[!b]
\includegraphics[width=8cm]{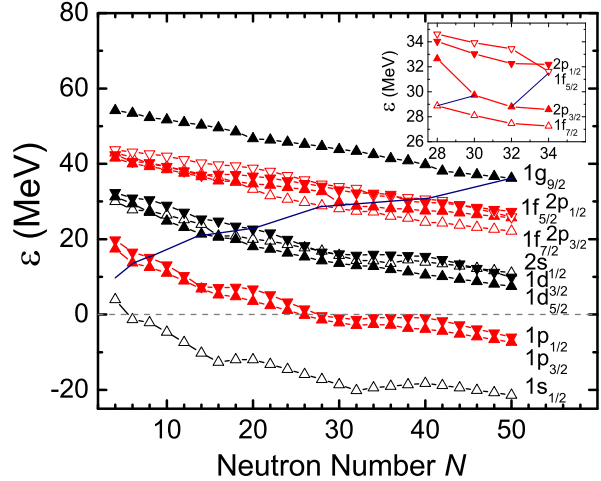}
\caption{Single-particle energies of $N$-neutron drops in a HO trap ($\hbar\omega = 10$ MeV) as a function of $N$ calculated by RBHF theory using the interaction Bonn A.
The blue line represents the Fermi surface.}
\label{fig10}
\end{figure}

In Fig.~\ref{fig10}, we show the single-particle energies of $N$-neutron drops in a HO trap ($\hbar\omega = 10$ MeV) as a function of $N$ calculated by RBHF theory using the interaction Bonn A. The blue line represents the Fermi surface. The filling approximation is used for open shell neutron drops.

Generally, the single-particle energies decrease as the number of neutron increases, because the potential becomes wider with increasing neutron number $N$.
The inset in Fig.~\ref{fig10} shows the details of $1f$ and $2p$ orbits in the region between $N = 28$ and  $N=34$. We observe how the traditional sub-shell closure at $N = 28$ disappears and a new closure at $N = 32$ appears in neutron drops.

\begin{figure}[!t]
\includegraphics[width=8cm]{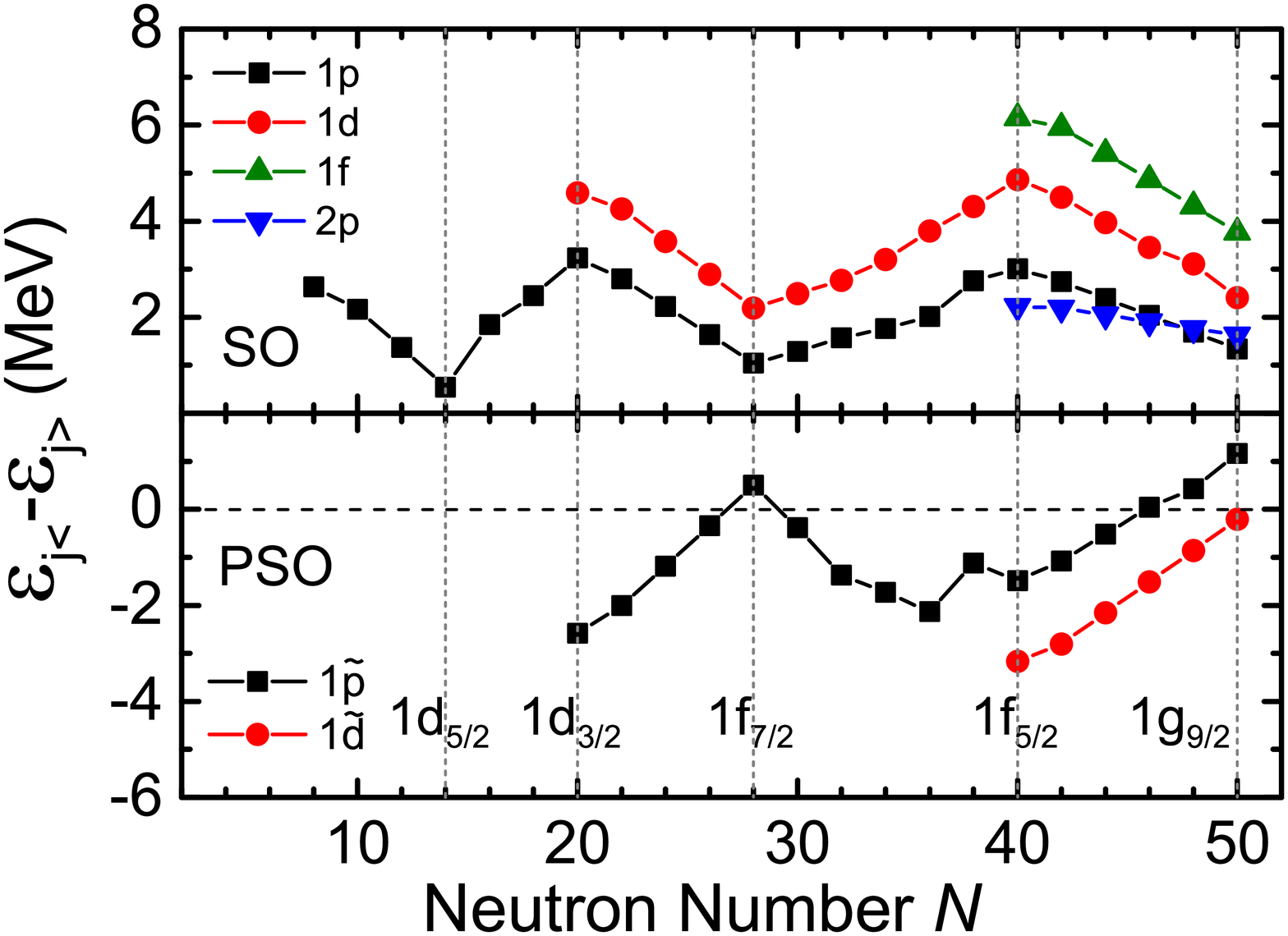}
\caption{Neutron spin-orbit and pseudospin-orbit splittings of $N$-neutron drops in a HO trap ($\hbar\omega = 10$ MeV) as a function of $N$ calculated by RBHF theory using the interaction Bonn A.}
\label{fig11}
\end{figure}

Another interesting phenomenon can be seen in the upper panel of Fig.~\ref{fig11}. It shows the evolution of spin-orbit (SO) splitting as the neutron number increases.
The SO splitting decreases as the next higher $j=j_>= l + 1/2$ orbit is filled and reaches a minimum when this orbit is fully occupied. As the number of neutron continues to increase, the $j=j_< = l - 1/2$ orbit begins to be occupied and the SO splitting increases.

A similar effect has been found by Otsuka et al. \cite{Otsuka2005_PRL95-232502}.
They explained it in terms of the monopole effect of the tensor force, which produces an attraction between a proton in a SO aligned orbit with $j=j_>=l + 1/2$ and a neutron in a SO anti-aligned orbit with $j'=j'_<=l' - 1/2$ and a repulsion between the same proton and a neutron in a SO aligned orbit with $j'=j'_>=l'+ 1/2$.

As discussed in the same paper~\cite{Otsuka2005_PRL95-232502}, a similar mechanism with smaller amplitude exists also for the tensor interaction between neutrons with T = 1. The behavior of the SO splitting in Fig.~\ref{fig11} has been explained in a similar way in Ref.~\cite{Shen2018} qualitatively.
Consider, for instance, the decreasing of the $1d$ SO splitting from $N=20$ to $N=28$.
Above $N=20$ the neutrons fill into the SO aligned orbit $1f_{7/2}$.
They show repulsion with the SO aligned $1d_{5/2}$ neutrons and attraction with the SO anti-aligned $1d_{3/2}$ neutrons.
This means by filling in neutrons into the $1f_{7/2}$ shell the $1d_{5/2}$ orbit is shifted upward and the $1d_{3/2}$ is shifted downward, reducing the $1d$ SO splitting more and more.
Above $N=28$ the neutrons fill into the SO anti-aligned states $2p_{1/2}$ and $1f_{5/2}$. They interact with the $1d$-neutrons in the opposite way and increase the $1d$-SO splitting.

\begin{figure}[!t]
\includegraphics[width=8cm]{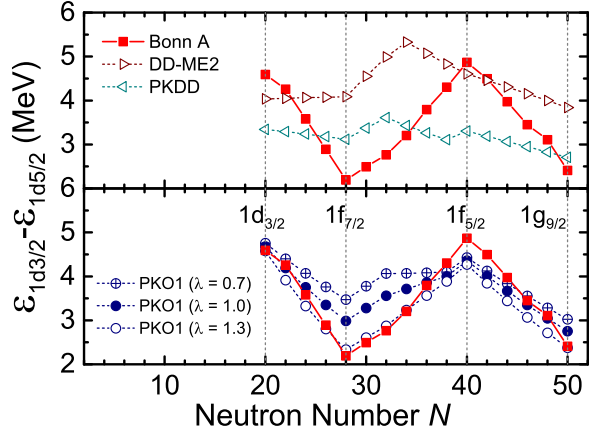}
\caption{Neutron spin-orbit for the 1$d$-orbit as a function of the neutron number $N$. Full RBHF calculations with the interaction Bonn A are compared with the two conventional density functionals PKDD and DD-ME2 without tensor contributions (upper panel) and with the RHF-functional PKO1, which contains a tensor contribution in the Fock term of the pion-exchange force (lower panel). Its strength $f_\pi$ is slightly varied (by a factor $\lambda$) as compared to the strength in PKO1.}
\label{fig12}
\end{figure}

In order to study, whether the characteristic variation of the SO splitting shown in the upper panle of Fig.~\ref{fig11} is indeed connected with the properties of the tensor force, the results of RBHF with Bonn A interaction have been compared with various relativistic density functionals, with and without tensor force \cite{Shen2018}. It has been found that the tensor force is the major reason for this pattern of the evlution of SO splittings. In
Fig.~\ref{fig12} our RBHF results for the 1d spin-obit splitting with results obtained with phenomenological density functionals from the literature.
Most of them, as for instance the functionals DD-ME2~\cite{Lalazissis2005_PRC71-024312} or PKDD~\cite{Long2004_PRC69-034319}, are based on relativistic Hartree calculations and do not include a tensor term, and, indeed, as shown in the upper panel of Fig.~\ref{fig12},  these functionals are not able to reproduce the specific pattern for the 1$d$ splitting.

On the other hand, the Hartree-Fock functionals PKO1~\cite{Long2006_PLB640-150} and PKA1~\cite{Long2007_PRC76-034314} include in the Fock term of the pion- and of the $\rho$-meson exchange forces tensor terms, PKO1 only for the pion and PKA1 for both the pion and for the $\rho$.
In the lower panel of Fig.~\ref{fig12} it is clearly seen that the SO splitting produced by the density functional PKO1 closely follows the pattern of our \emph{ab initio} RBHF calculations.
By changing the strength of the pion-exchange, i.e. by changing the  size of the corresponding tensor term it clearly seen that, the size of the tensor effect does significantly depend on the value of $\lambda$, where the cases of
$\lambda = 0.7,\,1.0$, and $1.3$ are shown in the figure. For $\lambda = 1$ we have the results of the density functional PKO1.
With $\lambda = 1.3$, the specific evolution pattern of the SO splitting generated by the \emph{ab initio} RBHF calculations can be nicely reproduced.
As in the case of the shell model calculations of the Otsuka et al.~\cite{Otsuka2005_PRL95-232502}, it is the tensor which causes the peculiar behavior of the SO splitting of the drops with increasing neutron number.
The pattern for the functional PKA1 is similar, therefore we did not present here.


The pseudospin-orbit (PSO) splitting~\cite{Arima1969_PLB30-517,Hecht1969_NPA137-129,Ginocchio1997_PRL78-436,Liang2013,
Shen2013,Liang2015_PR570-1} in the lower panel of Fig.~\ref{fig11} shows a similar pattern, but in opposite direction to the SO splitting. This can also be understood by the effect of the tensor force.
Taking the PSO splitting of $1\tilde{p}$ as an example, it is defined as $E_{\tilde{j}<} - E_{\tilde{j}>} = E_{2s1/2} - E_{1d3/2} = E_{j'>} - E_{j''<}$, with $l'$ the $s$ orbit and $l''$ the $d$ orbit \cite{Liang2015_PR570-1}.
As the $j_>$ orbit, for example $1f_{7/2}$, is being occupied, there will be a repulsion for $j_>' = 2s_{1/2}$ and an attraction for $j_<'' = 1d_{3/2}$.
Thus, the PSO splitting of $1\tilde{p}$ will increase and even become positive at $N = 28$, when $j_> = 1f_{7/2}$ is fully occupied. The situation becomes more complicated when the $2p$ orbit is being filled and the monopole effect of the tensor force is less prominent.

\section{Summary}\label{sec:sum}

We have studied neutron drops confined in an external field of oscillator shape using relativistic Brueckner-Hartree-Fock theory with the relativistic $NN$ interactions Bonn A, B, and C \cite{Machleidt1989_ANP19-189}. The results are compared with other nonrelativistic \emph{ab initio} and with relativistic density functional calculations.

First, we checked the convergence of RBHF calculations for neutron drops with $N = 8,\,20,\,28,\,50$ with respect to the single-particle energy cut-off $\varepsilon_{\rm cut}$ and found good convergence at $900$ MeV. This energy cut-off is smaller than the cut-off in finite nuclei at $1100$ MeV. This can be understood by the lack of the $T = 0$ tensor force in neutron drops. We also showed that this energy cut-off does not depend on the number of particles.

We calculated $N = 4$ to $28$ neutron drops with the interactions Bonn A, B, and C and found similar results for these three interactions. We compared for Bonn A our RBHF results for $N = 4$ to $N=50$ neutrons with other \emph{ab initio} calculations and with various relativistic density functionals.
The harmonic oscillator magic numbers $8,\,20,\,40$ show up in all the selected results, but the sub-shell closures at $N = 28$ and $N = 32$ strongly depend on the interactions.
There is little sign of a sub-shell closure at $N = 28$ for AV8' + UIX, but no sign for the other interactions.
For $N = 32$, AV8' + IL7 shows a clear sub-shell closure, while the sub-shell closure for RBHF with Bonn A and for AV8' is smaller but still significant. On the other hand, relativistic density functionals show only the HO magic numbers.

We also studied the radii of neutron drops in a HO trap. With increasing $N$ they follow closely the $N^{1/6}$ rule, which can be derived for non-interacting neutrons in Thomas-Fermi approximation. While the energies of RBHF with Bonn A are similar to those of the JISP16 interaction, the radii of RBHF are smaller. On the other hand, the radii calculated by various relativistic density functionals are all larger than RBHF with Bonn A, even though the energies found in RBHF are among these obtained with the density functionals. However, the smaller radii given by RBHF are in good agreement with pseudo-data derived from the experimental neutron skin thickness of $^{48}$Ca and $^{208}$Pb. These pseudo-data are derived from the strong linear correlation found in Ref.~\cite{ZHAO-PW2016_PRC94-041302} between the radius of a fixed neutron drop and the neutron skin thickness of a specific nucleus for various nuclear density functionals.
In particular we have calculated the neutron skin thickness of $^{48}$Ca by RBHF with Bonn A and the value is consistent with recent experimental datum \cite{Birkhan2017} and coupled-cluster calculations \cite{Hagen2015_NatP3529}.

We show the density distribution of neutron drops with $N = 8,\,20,\,32,\,40,\,50$ and find that the density gets saturated around $0.14 - 0.17$ fm$^{-3}$.
Similarly, we calculated the local equivalent single particle potentials for the $1s_{1/2}$ states and find also saturation for $N \geq 20$ at a potential depth of around $-40$ MeV.
These results depend on the strength of the external HO field. We have used $\hbar\omega = 10$ MeV, and changing the strength will change the saturation properties.

Finally we studied the evolution of the single-particle energies as a function of $N$.
The disappearance of a sub-shell closures at $N = 28$ and appearance at $N = 32$ can be seen clearly.
We also find that the evolution of the spin-orbit and the pseudospin-orbit splittings show a interesting pattern, which can be explained in a similar way by the tensor force as it has been done in nuclei in Ref. \cite{Otsuka2005_PRL95-232502}.

The results of RBHF show many interesting features and can provide important information for future density functionals, especially in the area of neutron-rich exotic nuclei.
To name a few for future guidelines,
\begin{enumerate}
  \item It is evident from the results on the spin-orbit splitting, that we have to introduce a tensor term. We need a further study on several tensor terms (zero-range, pion-like, rho-like) to find out which of them is the most appropriate.
  \item One could adjust the parameters of future relativistic density functionals not only to the conventional data on nuclear matter and finite nuclei, but also to the matrix elements of the $G$-matrix in specific nuclei.
  \item In a more systematic way, one could decompose the $G$-matrix into the different relativistic channels and to study, which of them are important for specific types of nuclei.
  \item Applying external fields of various types and studying their influence on the RBHF-results will allow to model the corresponding relativistic density functionals. An example would be the application of an external magnetic field in order to study the time-odd parts of the functionals. Another example would be the solution of half-infinite nuclear matter in the RBHF framework for the study of the surface properties of the functionals.
\end{enumerate}
Of course, these are only examples and details have to be investigated in future. However, it is evident, that the knowledge of the G-matrix in finite systems opens a completely new field of investigations to improve the functionals.

So far, there is only one relativistic nucleon-nucleon force, the Bonn potential. With the recent progress in covariant chiral interactions \cite{RenXL2018,LiKW2018} it will be also interesting to study the neutron drops using RBHF theory with covariant chiral interactions.

\section*{ACKNOWLEDGMENTS}

We thank Pengwei Zhao for discussions and providing his results.
This work was partly supported by the Major State 973 Program of China No.~2013CB834400, Natural Science Foundation of China under Grants No.~11335002, No.~11375015, and No.~11621131001,  the Overseas Distinguished Professor Project from Ministry of Education No.
MS2010BJDX001, the Research Fund for the Doctoral Program of Higher Education under Grant No.~20110001110087, and the DFG (Germany) cluster of excellence \textquotedblleft Origin and Structure of the Universe\textquotedblright\ (www.universe-cluster.de).
HL would like to thank the RIKEN iTHES project and iTHEMS program.

\bibliographystyle{apsrev4-1}

\end{CJK}
\end{document}